\documentclass[preprint,onecolumn,nofootinbib]{revtex4}

\usepackage[colorlinks=true,
            linkcolor=blue,
            urlcolor=blue,
            citecolor=red,
            pdfstartview=FitV,
            bookmarksopen=true]{hyperref}
\usepackage{amsmath, amsfonts, amssymb}
\usepackage{graphicx}
\usepackage{xcolor}
\usepackage{tikz}
\usepackage{autobreak}
\usepackage{enumerate}
\usepackage{dcolumn}
\usepackage{ulem}

\allowdisplaybreaks

\begin{document} 

\title{Quantum criticality and mixed-state entanglement in holographic superconductor--insulator transitions}

\author{Zhe Yang $^{1}$}
\email{yzar55@stu2021.jnu.edu.cn}
\author{Fang-Jing Cheng $^{3,4,5}$}
\email{fjcheng@mail.bnu.edu.cn}
\author{Guoyang Fu $^{1}$}
\email{FuguoyangEDU@163.com}
\author{Yi Ling $^{4,5}$}
\email{lingy@ihep.ac.cn}
\author{Peng Liu $^{2}$}
\email{phylp@email.jnu.edu.cn}
\thanks{Corresponding author}
\author{Jian-Pin Wu $^{1}$}
\email{jianpinwu@yzu.edu.cn}
\thanks{Corresponding author}

\affiliation{
  $^{1}$ \mbox{Center for Gravitation and Cosmology, College of Physical Science and Technology}, \mbox{Yangzhou University, Yangzhou 225009, China}\\  
  $^{2}$ \mbox{Department of Physics and Siyuan Laboratory,} Jinan University, Guangzhou 510632, China\\
  $^{3}$ \mbox{School of Physics and Astronomy, Beijing Normal University, Beijing 100875, China}\\
  $^{4}$ \mbox{Institute of High Energy Physics, Chinese Academy of Sciences, Beijing 100049, China}\\
  $^{5}$ \mbox{School of Physics, University of Chinese Academy of Sciences, Beijing 100049, China}\\
}

\begin{abstract}
  We study quantum criticality in a holographic Einstein--Maxwell--Dilaton--Axion (EMDA) p-wave superconductor exhibiting a superconductor--insulator transition (SIT). By tracking the superconducting energy gap, we show that approaching the quantum critical point (QCP) closes the gap and induces incipient insulating features, indicating that enhanced quantum fluctuations suppress superconducting order and trigger the SIT. We suggest that this behavior occurs only when the condensate orientation is aligned with the direction of translational symmetry breaking. To probe the transition, we employ two holographic indicators: holographic entanglement entropy (HEE) and the entanglement wedge cross-section (EWCS), the latter being a mixed-state entanglement measure. In contrast to HEE, which for sufficiently large configuration is dominated by the thermal entropy and is therefore largely insensitive to entanglement along the temperature direction, EWCS displays pronounced critical scaling and provides a robust diagnostic of the quantum phase transition (QPT). We attribute this contrast to the fact that HEE at large scales is controlled by the infrared (IR) geometry, whereas EWCS is governed by deformations of the entire bulk. Our results establish EWCS as a robust probe of holographic quantum criticality in mixed states.
\end{abstract}
\maketitle
\tableofcontents
\section{Introduction}

Quantum criticality governs continuous phase transitions in heavy-fermion compounds, strange metals, and correlated Fermi systems~\cite{sachdev2011quantum,coleman2005quantum,gegenwart2008quantum,phillips2022stranger}. Near a quantum critical point (QCP), scale-invariant quantum fluctuations proliferate across wide energy and length scales, driving pronounced deviations from standard thermodynamics. These ideas increasingly interface with quantum information science and holographic duality, providing a unified lens on correlations and universality~\cite{sachdev1999quantum,chertkov2023characterizing}.

Quantum information measures offer powerful probes of critical behavior~\cite{osterloh2002scaling,amico2008entanglement}. Entanglement entropy (EE) is a standard diagnostic of quantum phase transitions (QPTs) for pure states, exhibiting universal scaling in diverse settings~\cite{nielsen2010quantum,wilde2013quantum,hayashi2006quantum}. For mixed states, however, EE conflates classical and quantum correlations, rendering it inadequate at finite temperature. This limitation has motivated mixed-state entanglement quantifiers---such as entanglement of purification, reflected entropy, and logarithmic negativity---that disentangle classical from quantum components~\cite{vidal2002computable,plenio2005logarithmic,horodecki2009quantum}. Yet direct computation of these quantities in strongly correlated systems is notoriously challenging.

Holographic duality (AdS/CFT) offers a complementary approach by mapping strongly coupled many-body systems to classical gravitational theories~\cite{hooft1993dimensional,susskind1995world,maldacena1999large,witten1998anti,zaanen2015holographic,baggioli2019applied,ammon2015gauge,natsuume2015ads}. In this framework, EE is computed geometrically via the Ryu--Takayanagi (RT) formula, and holographic entanglement entropy (HEE) has been widely applied to phase transitions~\cite{ryu2006holographic,cai2012holographic,peng2014holographic,ling2016holographic,ling2016characterization}. For mixed states, the entanglement wedge cross-section (EWCS) provides a geometric measure of correlation structure~\cite{umemoto2018entanglement1,umemoto2018entanglement2}, and has been conjectured to capture entanglement of purification, reflected entropy, and logarithmic negativity~\cite{dutta2021canonical,kudler2019entanglement,jokela2019notes,camargo2022balanced,vasli2023holographic}. Building on these tools, we investigate quantum criticality in holographic models using mixed-state entanglement measures.

Holographic superconductors provide effective models of high-$T_c$ superconductivity: s-wave superconductors emerge from charged scalar hair, p-wave from charged vector fields, and d-wave from massive spin-2 fields~\cite{hartnoll2008holographic,horowitz2011introduction,hartnoll2008building,cai2012holographic}. The superconducting phase is characterized by an energy gap that stabilizes the condensate. Experimentally, this superconducting gap is accessed by angle-resolved photoemission spectroscopy (ARPES)~\cite{giaever1960energy,hashimoto2014energy,damascelli2004probing,bansil1999importance,boschini2024time}, and we can compute the holographic fermionic spectral function~\cite{liu2011non,iqbal2012lectures,faulkner2010photoemission,faulkner2011emergent}.

In contrast to finite-temperature transitions, holographic QPTs are driven by deformations of the infrared (IR) fixed point at zero temperature, as exemplified by metal--insulator transitions~\cite{hartnoll2018holographic,donos2013interaction,donos2014holographic,donos2014novel}. Our motivation is to explore how symmetry breaking and IR deformations interplay within a single model. We study an Einstein-Maxwell-Dilaton-Axion (EMDA) p-wave superconductor with multiple control parameters, giving rise to both superconducting transitions and a superconductor--insulator transition (SIT). Since entanglement has been suggested to play a pivotal role in QPTs, we employ holographic quantum information probes to diagnose the transitions. In particular, we focus on the EWCS as a mixed-state entanglement measure, leveraging its demonstrated utility in related holographic models to reveal the critical structure and broaden our understanding of the underlying mechanisms. Notably, we find that the SIT arises only in the EMDA p-wave superconductor model, whereas no such transition is observed in the EMDA s-wave model. We attribute this qualitative difference to the interplay between axion-induced translation-symmetry breaking and p-wave condensation. The axion field implements an effective lattice deformation that can drive the normal state toward insulating behavior, and the p-wave order parameter appears more sensitive to this deformation, thereby making the SIT accessible only in the EMDA p-wave superconductor model.

In this paper, we systematically study the holographic EMDA p-wave superconductor and the holographic quantum information. Section \ref{sec:holosetup} presents the EMDA p-wave model, the phase diagram encompassing superconducting and superconducting–insulating transitions, and the superconducting energy gap. Section \ref{sec:qi} examines holographic quantum information across the critical point, including HEE and EWCS. Section \ref{sec:discu} summarizes our findings, and Appendix \ref{ap:eom} records the equations of motion.

\section{Holographic setup and phase diagram}\label{sec:holosetup}

We construct a p-wave holographic superconductor in an EMDA background. After presenting the model and ansatz, we map out the phase diagram---identifying normal, superconducting, and insulating regimes---and use the fermionic spectral function to diagnose the SIT via the superconducting energy gap.

\subsection{The holographic model}

Below the critical temperature $T_{c1}$, a holographic p-wave superconductor undergoes spontaneous $U(1)$ symmetry breaking, forming a vector order parameter~\cite{cai2014towards,nie2013competition,li2015entanglement,yang2023mixed,cai2015introduction}. In the EMDA model, zero-temperature phases are controlled by the emergent IR geometry; distinct IR solutions correspond to distinct quantum ground states~\cite{fu2022novel,gong2023diagnosing}. We couple a p-wave superconductor to an EMDA background to explore the interplay between symmetry breaking and IR deformations. The action is
\begin{equation}\label{eq:eom}
  \begin{aligned}
  &S=\frac{1}{2\kappa}\int d^4 x\sqrt{-g}\left(\mathcal{R}+\frac{1}{4}Z(\Psi)F^2+\mathcal{L}_\Psi+\mathcal{L}_\phi\right),\\
  &\mathcal{L}_\Psi=-\frac{3}{2}((\partial \Psi)^2+Y(\Psi)(\partial\chi^2))-V(\Psi),\\
  &\mathcal{L}_\phi=-\frac{1}{2}\rho_{\mu\nu}^\dagger \rho^{\mu\nu}-m^2\rho_\mu^\dagger\rho^\mu+iq\eta\rho_\mu\rho_ \nu^\dagger F^{\mu\nu},
  \end{aligned}
\end{equation}
where $\kappa^2=8\pi G$, $F=dA$ is the field strength of Maxwell field, $\Psi$ is the dilaton field, $\chi$ is the axion field, $\rho_\mu$ is the complex vector field with mass $m$ and charge $q$. The last term of $\mathcal{L}_\phi$ represents the coupling between the Maxwell field and the complex vector field; since we do not include a magnetic field, we set $\eta=0$. These fields are defined as follows: 
\begin{equation}
  \begin{aligned}
    &V(\Psi)=-6\text{Cosh}(\Psi),\quad Z(\Psi)=\text{Cosh}^{\gamma/3}(3\Psi),\quad Y(\Psi)=4\text{Sinh}^2(\Psi),\\
    &\rho_{\mu\nu}=\mathcal{D}_\mu\rho_\nu-\mathcal{D}_\nu\rho_\mu, \quad \mathcal{D}_\mu=\nabla_\mu-iq A_\mu.\\
  \end{aligned}
\end{equation}
Additionally, we consider only $\chi$ depends on a specific spatial direction along $x$, which introduces anisotropy in the background. The equation of motion (EOM) of the action can be read in Appendix~\ref{ap:eom}.

We can solve the EOM with the ansatz 
\begin{equation}
  \begin{aligned}
    &ds^2=\frac{1}{z^2}\left(-p(z)(1-z)U(z)dt^2+\frac{1}{p(z)(1-z)U(z)}dz^2+V_1(z)dx^2+V_2(z)dy^2\right),\\
    &A_\mu dx^\mu=\mu(1-z)a(z)dt,\quad \rho_\mu dx^\mu=\rho_x(z)dx,\quad\\
    &\Psi=z^{3-\Delta}\psi(z),\quad \chi=\hat{k} x.
  \end{aligned}
\end{equation}
The scalar field $\Psi$ has conformal dimension $\Delta=2$, and we set $p(z)=1+z+z^2-\frac{\mu^2 z^3}{4}$. Here $\mu$ denotes the chemical potential in the dual field theory, while $\hat{k}$ is the lattice wave number. The radial coordinate $z\in(0,1)$ parametrizes the bulk geometry, with $z=0$ corresponding to the AdS boundary and $z=1$ to the horizon. The background is characterized by six unknown functions $U(z)$, $V_1(z)$, $V_2(z)$, $a(z)$, $\rho_x(z)$, and $\psi(z)$, which are determined by solving the EOM subject to appropriate boundary conditions. Near the AdS boundary, $\rho_x$ admits the asymptotic expansion
\begin{equation}
\rho_x=\rho_{x_-}z^{\Delta_{-}}+\rho_{x+}z^{\Delta_+}+\cdots.
\end{equation}
We set the source term to zero, $\rho_{x_-}=0$, so that the condensate forms spontaneously. The condensate $\langle J_x\rangle$ is then obtained by extracting the coefficient $\rho_{x_+}$. To preserve the asymptotic $AdS_4$ structure at the boundary, we impose the following boundary conditions
\begin{equation}
  U(0)=1,\quad V_1(0)=1,\quad V_2(0)=1,\quad a(0)=1,\quad \psi(0)=\hat{\lambda},\quad \rho_x(0)=0.
\end{equation}
Here, $\hat{\lambda}$ denotes the source term for the dilaton operator and thus parametrizes the amplitude of the lattice deformation. The superconductor is signaled by the emergence of a non-vanishing condensate $\langle J_x\rangle$, such that the dual vector operator develops a finite vacuum expectation value, leading to spontaneous breaking of the $U(1)$ symmetry. In this setup, the Hawking temperature is  $\hat{T}=\frac{12-\mu^2}{16\pi}$ and the system is invariant under the following rescaling,
\begin{equation}
  \begin{aligned}
    &(t,x,y)\to\alpha^{-1}(t,x,y),\quad (U,V_1,V_2)\to\alpha^2(U,V_1,V_2),\quad \mu\to\alpha\mu\\
    &\hat{T}\to\alpha\hat{T},\quad \rho_{x_+}\to\alpha^{\Delta_{+}+1}\rho_{x_+},\quad \hat{k}\to \alpha \hat{k},\quad \hat{\lambda}\to\alpha\hat{\lambda}.
  \end{aligned}
\end{equation}
In this paper, we adopt chemical potential $\mu$ as the scaling unit, which is equivalent to treating the dual system as a field theory described by the grand canonical ensemble. Therefore, we have three dimensionless parameters $\{T,k,\lambda\}=\{\hat{T}/\mu,\hat{k}/\mu,\hat{\lambda}/\mu\}$.
\subsection{Phase diagram}
\begin{figure}
  \centering
  \includegraphics[width=0.4\textwidth]{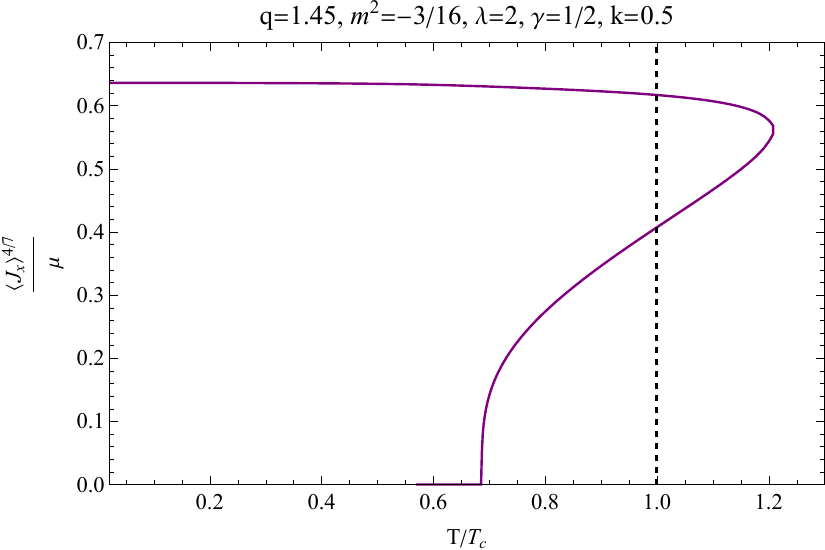}\qquad
  \includegraphics[width=0.4\textwidth]{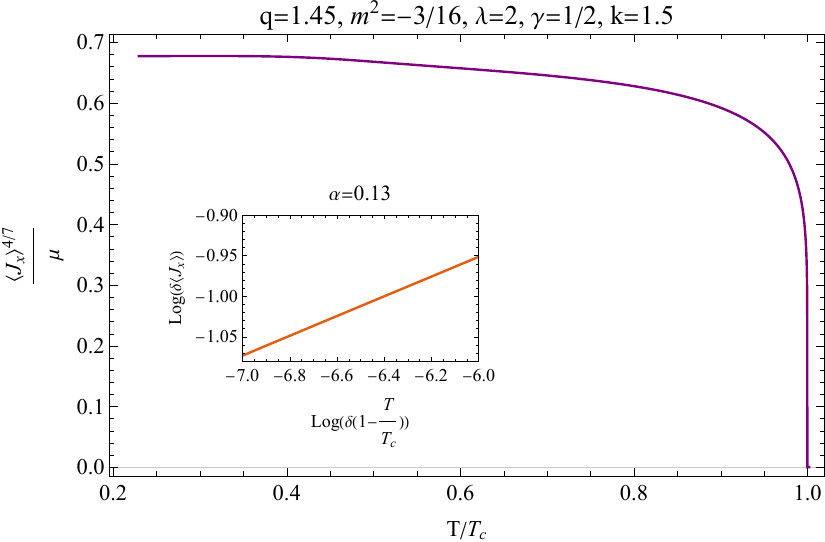}\\
  \includegraphics[width=0.4\textwidth]{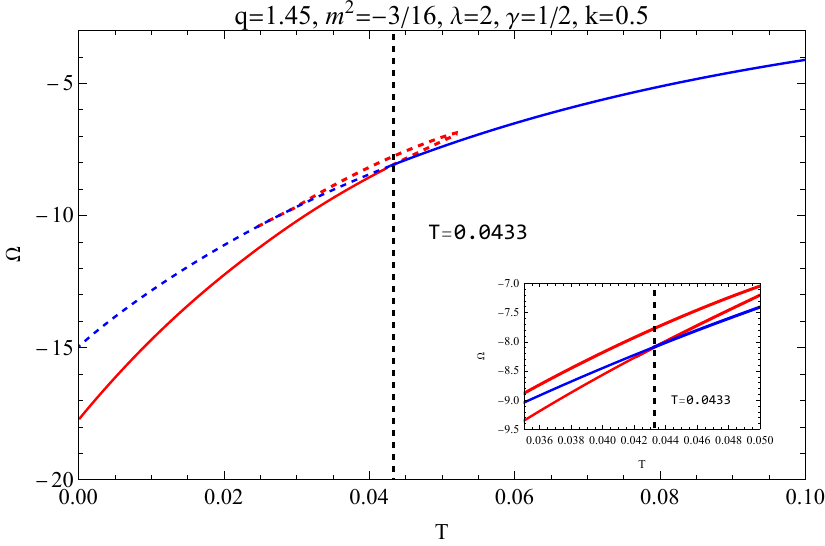}\qquad
  \includegraphics[width=0.4\textwidth]{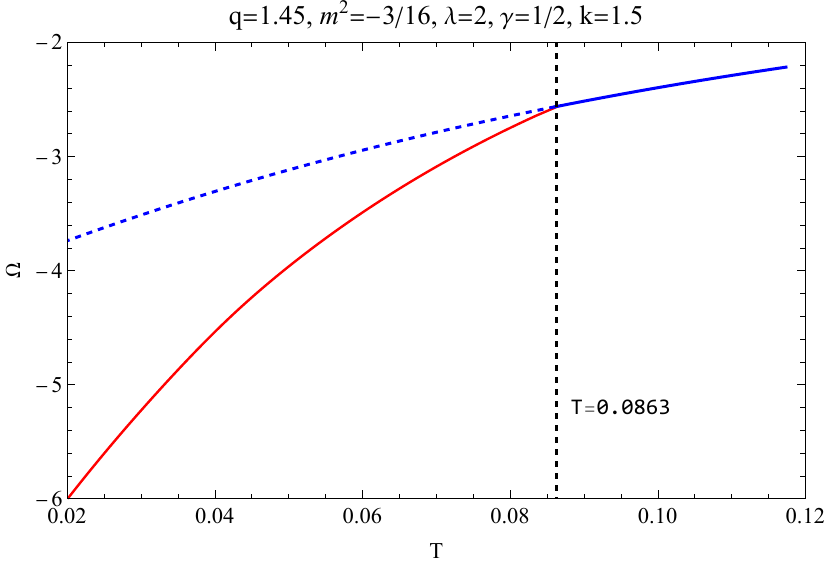}
  \caption{Both first-order and second-order superconducting phase transitions occur when the temperature is lowered below the critical temperature $T_c$. The upper panels show the condensate $\langle J_x\rangle$ as a function of the temperature $T/T_c$, while the lower panels show the free energy density $\Omega$ as a function of $T$. The red lines represents the superconducting states and the blue lines represents the normal states. The left panels correspond to the case $k=0.5$, where the system undergoes a first-order phase transition at $T_c=0.0433$ (black dashed line). The inset figure shows the detail of the $\Omega$ near the transition point. The thermodynamically preferred phase is identified by the branch with lower $\Omega$, and the transition occurs where the two branches exchange dominance. The right panels correspond to the case $k=1.5$, where a second-order phase transition takes place at $T_c=0.0863$. The inset figure shows the scaling behavior of the order parameter, and the critical exponent is $\alpha=0.13$.}
  \label{fig:suppd}
\end{figure}

Figure~\ref{fig:suppd} shows the condensate $\langle J_x\rangle$ versus temperature $T$. For $T > T_c$, $\langle J_x\rangle = 0$ (normal phase); for $T < T_c$, a nonzero condensate forms (superconducting phase). In this model, both first-order and second-order phase transitions can occur. In Fig.~\ref{fig:suppd}, we also present the free energy density associated with the phase transition. Since we consider only black brane solutions in this work, the total entropy is divergent due to the infinite spatial volume of the planar geometry. Therefore, throughout this paper we work with the free energy density, defined as $\Omega=\tilde{\Omega}/V_2$, where $\tilde{\Omega}$ denotes the free energy and $V_2$ is the two-dimensional spatial volume of the boundary theory. The free energy density is given by
\begin{equation}
\Omega=M-Ts-\mu \tilde{q},
\end{equation}
where $\tilde{q}$ is the charge density, which can be extracted from the subleading term in the asymptotic expansion of $A_t$. Here, $M$ denotes the ADM mass of the black brane and $s$ is the entropy density. The free energy provides a useful probe of the superconducting phase transition. For the case $k=0.5$, the free energy exhibits a swallow-tail structure, indicating a first-order phase transition at $T=0.0433$. By contrast, for $k=1.5$, the free energy of the superconducting phase is always lower than that of the normal phase. In this case, the first derivative of the free energy is continuous, while the second derivative is discontinuous, signaling a second-order phase transition. From the free energy, one can further determine the thermodynamic stability of the superconducting phase. In general, lowering the temperature enhances the thermodynamic stability of the superconducting state. It should also be emphasized that the restriction to a nonvanishing $\rho_x$ is part of the standard $p$-wave ansatz used in the present class of models \cite{cai2014towards,Cai:2013aca,cai2012holographic}. This follows directly from the ansatz $(0,0,\rho_x(z),0)$ for the vector field, under which all other components vanish identically. Substituting this ansatz into the equations of motion, one finds that only $\rho_x$ appears, implying that the condensate is restricted to the $x$ direction in the present setup.

\begin{figure}
  \centering
  \includegraphics[width=0.45\textwidth]{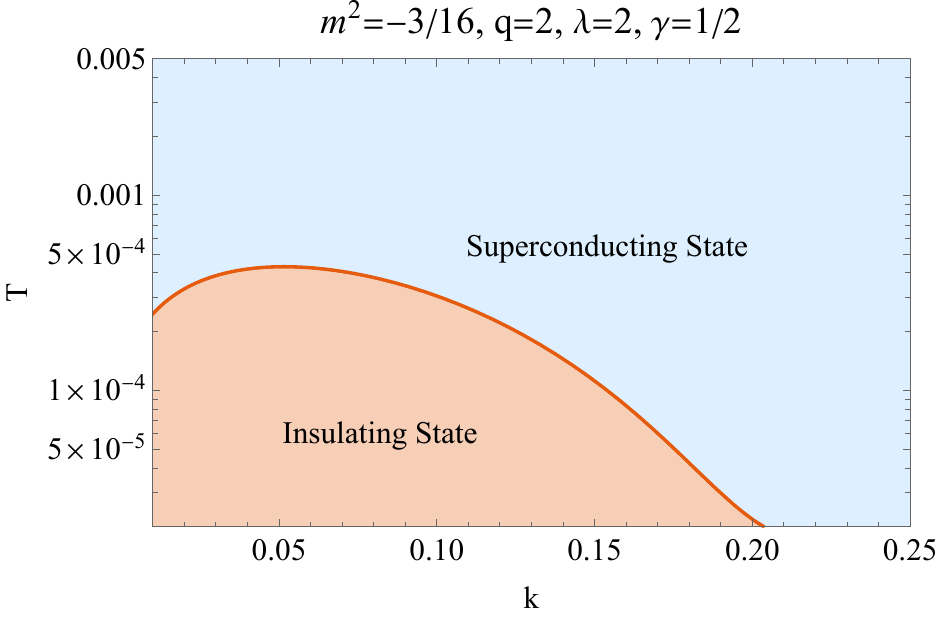}\qquad
  \includegraphics[width=0.45\textwidth]{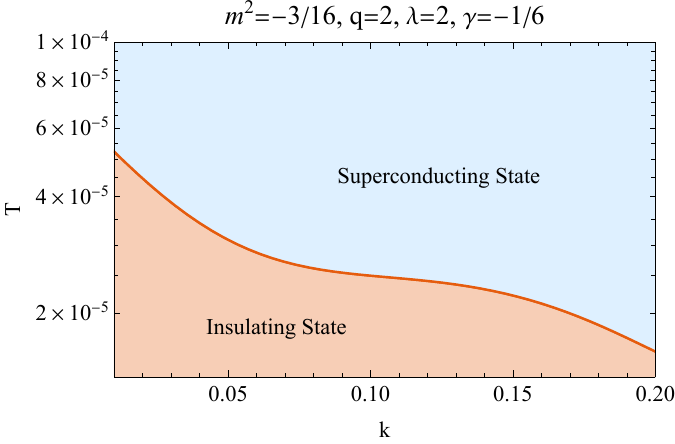}
  \caption{Phase diagrams in the $(T,k)$ plane showing the SIT as a zero-temperature QPT. The superconducting and insulating phases are separated by a critical line.}
  \label{fig:qpd}
\end{figure}

 The holographic EMDA p-wave superconductor, however, exhibits a novel phenomenon: superconductivity at finite temperature can be destabilized by proximity to a zero-temperature QCP. In the near zero-temperature regime, quantum fluctuations cannot be neglected and can suppress the superconducting condensate close to the QCP, leading to novel critical behaviour such as the SIT \cite{dubi2007nature,goldman1998superconductor,gantmakher2010superconductor,bollinger2011superconductor}. Our EMDA p-wave model exhibits analogous SIT behavior. As shown in figure~\ref{fig:qpd}, the phase diagram indicates that the relevant criticality is restricted to the $T\to 0$ limit, and the QCP shifts with both parameters $k$ and $\gamma$. For $\gamma = 1/2$, the $k$-dependence of the QCP is non-monotonic, exhibiting an “island” feature in the small-$k$ regime. By contrast, for $\gamma = -1/6$, the QCP decreases monotonically with increasing $k$, and the island region disappears. Importantly, the QPT is generically weakened in the large-$k$ regime, indicating that the SIT is favoured at small-$k$ regime and suppressed as $k$ grows. A natural explanation is that the EMDA model already tends toward an insulating transition in the small-$k$ regime, thereby predisposing the coupled system to exhibit SIT criticality there.

For the superconducting and insulating phases, the system flows to two distinct IR fixed points: one is characterized by a hyperscaling-violating geometry, while the other exhibits a novel scaling geometry. The instability of an IR fixed point generally signals the occurrence of a QPT in the zero-temperature limit. To investigate the IR fixed points more effectively, we employ the butterfly velocity $\nu_B$ as a diagnostic probe. The butterfly velocity characterizes the dynamical spread of quantum information and captures the propagation of chaos. Along the $x$ direction, it is given by  
\begin{equation}
  \label{eq:vb}
  \nu_B=\left. \sqrt{\frac{-2\pi T\mu V_2}{V_2(V_1'-2V_1)+V_1(V_2'-2V_2)}}\right|_{z=1}.
\end{equation}

It is straightforward to see that the butterfly velocity depends only on the IR geometry. Therefore, in the zero-temperature limit it exhibits the scaling behavior $\nu_B \sim T^{\alpha}$. As a result, $\nu_B$ provides a useful probe of the IR fixed point. In particular, one can extract the scaling exponent $\alpha$ by evaluating $T\nu_B'/\nu_B$, which distinguishes different IR fixed points.

For the insulating phase, the IR fixed-point solution can be written as  
\begin{equation}
  \label{eq:irsol}
  g_{tt}=g_{zz}^{-1}\sim z^{-u_1},\quad g_{xx}\sim z^{-v_1},\quad g_{yy}\sim z^{-v_2},\quad a\sim z^{-a_1},\quad e^\phi\sim z^{-\psi_1},
\end{equation}
with coefficients
\begin{equation}
  \label{eq:ircoef}
  \begin{aligned}
  &u_1=\frac{2(\gamma^2+3\gamma+10)}{\gamma^2+4\gamma+11},\quad v_1=\frac{-2(\gamma+1)}{\gamma^2+4\gamma+11},\quad v_2=\frac{2(\gamma+1)(\gamma+2)}{\gamma^2+4\gamma+11},\\
  & a_1=\frac{2(\gamma^2+2\gamma+5)}{\gamma^2+4\gamma+11},\quad \psi_1=-\frac{2(\gamma+1)}{\gamma^2+4\gamma+11}.
  \end{aligned}
\end{equation}
This demonstrates that the IR geometry of the insulating phase is hyperscaling violating. Substituting Eqs.~\eqref{eq:vb}, \eqref{eq:irsol}, and \eqref{eq:ircoef}, we obtain the scaling behavior of the butterfly velocity:
\begin{equation}
  \nu_B\sim T^{\frac{11+4\gamma+\gamma^2}{9+2\gamma+\gamma^2}}.
  \label{eq:irinsulator}
\end{equation}

\begin{figure}
  \centering
  \includegraphics[width=0.45\textwidth]{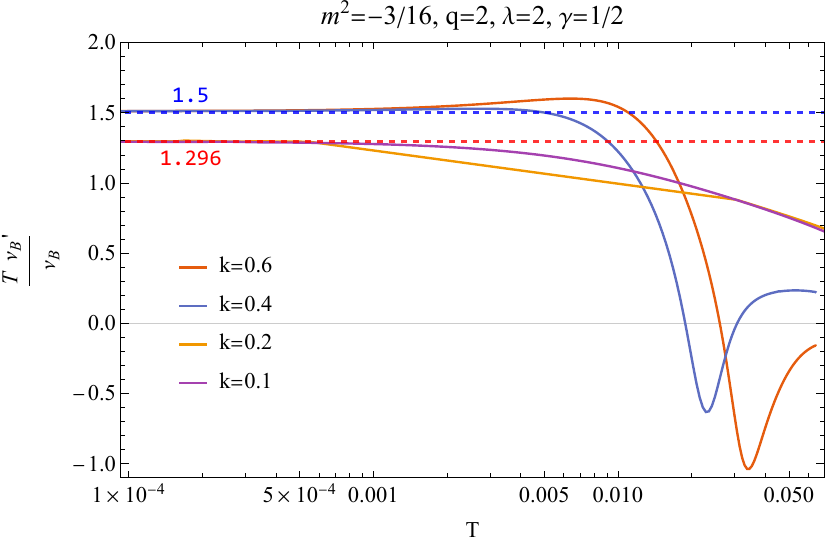}\qquad
  \includegraphics[width=0.45\textwidth]{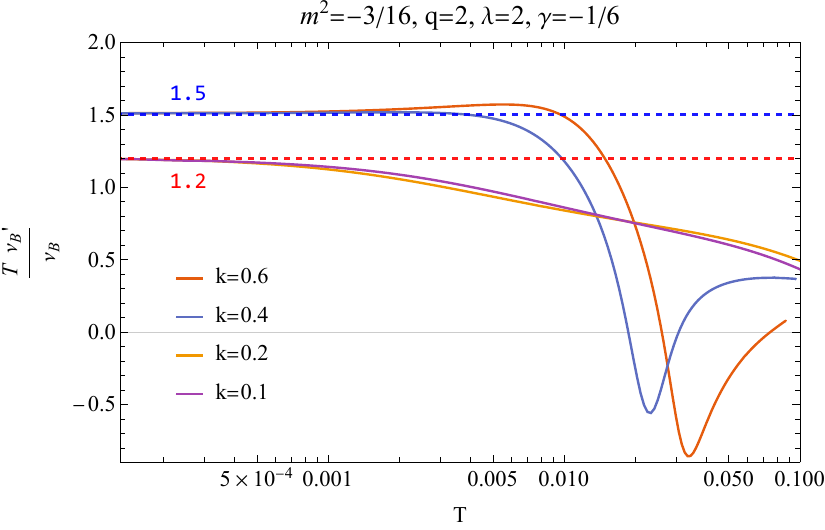}
  \caption{The $T \nu_B'/\nu_B$ as a function of temperature $T$, used to characterize the IR fixed points of different phases. The solid curves show the numerical results for the temperature dependence. The blue dashed line corresponds to the superconducting phase, for which the scaling exponent is $\alpha=1.5$. The red dashed line corresponds to the insulating phase, which scaling behavior is consistent with $\alpha=\frac{11+4\gamma+\gamma^2}{9+2\gamma+\gamma^2}$.}
  \label{fig:irfix}
  \end{figure}

As shown numerically in Fig.~\ref{fig:irfix}, the insulating phase exhibits the scaling exponent $\alpha=1.296$ for $\gamma=-1/2$ and $\alpha=1.2$ for $\gamma=-1/6$, in good agreement with Eq.~\eqref{eq:irinsulator}. More importantly, in contrast to the insulating phase, Fig.~\ref{fig:irfix} shows that the superconducting phase has a fixed exponent $\alpha=1.5$, independent of $\gamma$. This strongly suggests that the superconducting phase is governed by a novel IR fixed point determined by the p-wave condensate, which is fundamentally different from that of the insulating phase. The distinct IR fixed points associated with these two phases imply different ground states. Moreover, the instability of the IR fixed point provides the mechanism for the quantum phase transition.
However, although such instability can indicate a transition between two phases, it provides only limited information about the underlying critical behavior. To uncover the origin of the phase transition and identify definitive signatures of the QCP, we instead focus on the fermionic spectral function, whose frequency--momentum structure directly encodes the low-energy excitations across criticality.

\section{Fermionic spectral function and the superconducting gap}
In this section, we explore the spectral function of the EMDA p-wave superconductor model to probe the SIT. We compute the fermionic spectral function, which is accessible experimentally via ARPES~\cite{liu2011non,lee2009non,donos2014holographicfermi,ling2014holographic,ling2015holographic,wu2019dynamical}. We consider a probe fermion with dipole coupling:
\begin{equation} \label{eq:dirac1}
  S_D=i\int d^4 x \sqrt{-g}\bar{\zeta}(\Delta^a\mathcal{D}_a-ip(\rho_x)F \mkern -11.5 mu /)\zeta.
\end{equation}
In the action, the covariant derivative is defined as $\mathcal{D}_a=\partial_a+\frac{1}{4}(\omega_{\mu\nu})_a\Gamma^{\mu\nu}-iqA_a$, where $(\omega_{\mu\nu})_a$ denotes the spin connection, $A_a$ is the gauge potential, and $q$ is the fermion charge. We employ the standard notation $\Gamma^{\mu\nu}\equiv \tfrac{1}{2}[\Gamma^\mu,\Gamma^\nu]$, and introduce the dipole operator $F\mkern-11.5mu/ \equiv \frac{1}{2}\Gamma^{\mu\nu}(e_\mu)^a (e_\nu)^b F_{ab}$, with the orthonormal vielbein basis $(e_\mu)^a$. The function $p(\rho_x)$ depends on the condensed field $\rho_x$ and parametrizes the effective strength of the dipole coupling. In this paper, we take $p(\rho_x)=\tau\,\rho_x$, where $\tau$ is a constant parameter that controls the opening of the energy gap. Varying the action in Eq. \eqref{eq:dirac1} then yields
\begin{equation}
  \Gamma^a\mathcal{D}_a\zeta-m_\zeta \zeta-ip(\rho_x)F \mkern -11.5 mu /\zeta=0.
\end{equation}
To cancel off the spin connection, we make the redefinition of $\zeta=(g_{tt}g_{xx}g_{yy})^{-\frac{1}{4}}\mathcal{F}$ and consider the Fourier expansion $\mathcal{F}=\int\frac{d\omega dk_x dk_y}{2\pi}F(z,\textbf{k})e^{-i\omega t+ik_x x+ik_y y}$, so we have 
\begin{equation}\label{eq:dirac2}
  \begin{aligned}
  &-\frac{1}{\sqrt{g_{zz}}}\Gamma^3\partial_z F(z,\textbf{k})+\frac{1}{\sqrt{g_{tt}}}\Gamma^0(-i\omega-iqA_t)F(z,\textbf{k})+\frac{1}{\sqrt{g_{xx}}}\Gamma^1 ik_x F(z,\textbf{k})+\frac{1}{\sqrt{g_{yy}}}\Gamma^2i k_yF(z,\textbf{k})\\
  &-m_\zeta F(z,\textbf{k})+\frac{ip(\rho_x)}{\sqrt{g_{zz}g_{tt}}}\Gamma^3\Gamma^0\partial_z A_t F(z,\textbf{k})=0,
  \end{aligned}
\end{equation}
where $\textbf{k}=(\omega,k_x,k_y)$. We choose the following gamma matrices
\begin{equation}
  \begin{aligned}
  &\Gamma^3=\left(
  \begin{matrix}
    -\sigma^3 & 0\\
    0 & -\sigma^3
  \end{matrix}\right),\qquad
  \Gamma^0=\left(
  \begin{matrix}
    i\sigma^1 & 0\\
    0 & i\sigma^1
  \end{matrix}\right),\\
  &\Gamma^1=\left(
  \begin{matrix}
    -\sigma^2 & 0\\
    0 & \sigma^2
  \end{matrix}\right),\qquad \quad
  \Gamma^2=\left(
  \begin{matrix}
    0 & \sigma^2\\
    \sigma^2 & 0
  \end{matrix}\right),
\end{aligned}
\end{equation}
where $\sigma^i$ is Pauli matrix and we split spinor into two 2-component spinors $F=(F_1,F_2)^T$ and the decomposition $F_\alpha=(\mathcal{A}_\alpha,\mathcal{B}_\alpha)^T$. The Dirac equation can deduced from Eq.\eqref{eq:dirac2},
\begin{equation}
  \begin{aligned}
    &\left(\frac{1}{\sqrt{g_{zz}}}\mp m_\zeta\right)
    \left(\begin{matrix}
    \mathcal{A}_{1}\\
    \mathcal{B}_{1}
    \end{matrix}\right)\pm(\omega+q A_t)\frac{1}{\sqrt{g_{tt}}}
    \left(\begin{matrix}
      \mathcal{B}_1\\
      \mathcal{A}_1
    \end{matrix}\right)+\frac{p(\rho_x)}{\sqrt{g_{zz}g_{tt}}}\partial_z A_t 
    \left(\begin{matrix}
      \mathcal{B}_1\\
      \mathcal{A}_1
    \end{matrix}\right)\\
    &-\frac{k_x}{\sqrt{g_{xx}}}
    \left(
      \begin{matrix}
        \mathcal{B}_1\\
        \mathcal{A}_1
      \end{matrix}
    \right)+\frac{k_y}{\sqrt{g_{yy}}}
    \left(
      \begin{matrix}
        \mathcal{B}_2\\
        \mathcal{A}_2
      \end{matrix}
    \right)=0,\\
    &\left(\frac{1}{\sqrt{g_{zz}}}\mp m_\zeta\right)
    \left(\begin{matrix}
    \mathcal{A}_{2}\\
    \mathcal{B}_{2}
    \end{matrix}\right)\pm(\omega+q A_t)\frac{1}{\sqrt{g_{tt}}}
    \left(\begin{matrix}
      \mathcal{B}_2\\
      \mathcal{A}_2
    \end{matrix}\right)+\frac{p(\rho_x)}{\sqrt{g_{zz}g_{tt}}}\partial_z A_t 
    \left(\begin{matrix}
      \mathcal{B}_2\\
      \mathcal{A}_2
    \end{matrix}\right)\\
    &+\frac{k_x}{\sqrt{g_{xx}}}
    \left(
      \begin{matrix}
        \mathcal{B}_2\\
        \mathcal{A}_2
      \end{matrix}
    \right)+\frac{k_y}{\sqrt{g_{yy}}}
    \left(
      \begin{matrix}
        \mathcal{B}_1\\
        \mathcal{A}_1
      \end{matrix}
    \right)=0.
  \end{aligned}
\end{equation}
We need to impose the following independent ingoing boundary condition at the horizon to solve the Dirac equation, 
\begin{equation}
  \left(
    \begin{matrix}
      \mathcal{A}_\alpha\\
      \mathcal{B}_\alpha
    \end{matrix}
  \right)=c_\alpha
  \left(
    \begin{matrix}
      1\\
      -i 
    \end{matrix}
  \right)(1-z)^{-\frac{i\omega}{4\pi T}}.
\end{equation}
The boundary expansion of the Dirac field reads 
\begin{equation}
  \left(
    \begin{matrix}
      \mathcal{A}_\alpha\\
      \mathcal{B}_\alpha
    \end{matrix}
  \right)\approx a_\alpha z^{m_\zeta}
  \left(
    \begin{matrix}
      1\\
      0
    \end{matrix}
  \right)+b_\alpha z^{-m_\zeta}
  \left(
    \begin{matrix}
      0\\
      1
    \end{matrix}
  \right).
\end{equation}
From the holographic dictionary, the retarded Green function is obtained holographically
\begin{equation}
  a_\alpha=G_{\alpha{\alpha'}}b_{\alpha'}.
\end{equation}
We characterize the single-particle excitations through the spectral function, $A(\omega,\mathbf{k})\sim \text{Im}(\text{Tr}G_{\alpha{\alpha'}})$. As a momentum-resolved measure of electronic spectral weight, $A(\omega,\mathbf{k})$ distils the essential low-energy physics: coherent quasiparticle peaks delineate metallic band dispersions, while their gapping provides an immediate fingerprint of superconducting pairing~\cite{giaever1960energy,hashimoto2014energy,damascelli2004probing,bansil1999importance,boschini2024time}. In ARPES, the recorded intensity as a function of binding energy and crystal momentum approximates the spectral function, allowing direct comparison between holographic predictions and experiments~\cite{faulkner2010photoemission,damascelli2004probing,kordyuk2015pseudogap,hashimoto2014energy,boschini2024time,schmitt2008transient,souma2003origin,gao2024arpes,hwang2007high}.

\begin{figure}
  \centering
  \includegraphics[width=1\textwidth]{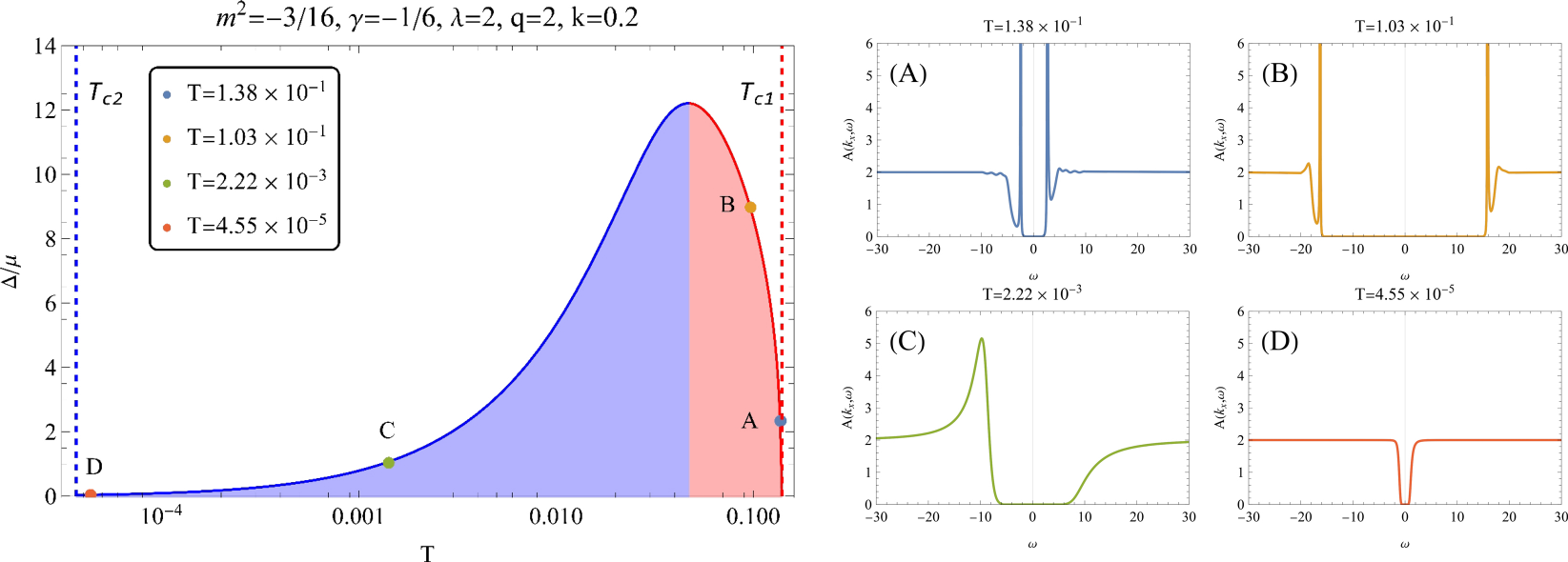}
  \caption{Temperature dependence of the superconducting gap $\Delta(T)$. A finite gap emerges for $T<T_{c1}=0.14$ and is driven to zero upon further cooling at $T_{c2}=1.5\times10^{-5}$, signaling the gap is closed. The red shaded region denotes the parameter region with the open superconducting gap, whereas the blue region corresponds to the gap-closed phase. Momentum-resolved spectral functions $A(\omega_x,k)$ at representative temperatures are presented in panels (A–D). As $T$ is lowered, the superconducting coherence features evolve markedly, consistent with a superconducting–insulator transition.}
  \label{fig:qpt}
\end{figure}

We focus on $k_y=0$ and define the superconducting gap $\Delta$ via the vanishing of $A(k_x,\omega)$. Figure~\ref{fig:qpt} summarizes the temperature dependence. Upon cooling through $T_{c1}$, the system enters the superconducting phase and the gap opens continuously. Strikingly, the gap does not grow monotonically---it first increases and reaches a maximum at intermediate temperatures, then decreases and closes at $T_{c2}$ in the zero-temperature limit. The absence of thermal criticality in this regime indicates that the transition at $T_{c2}$ is a QPT governed by quantum fluctuations. The right part of figure~\ref{fig:qpt} displays $A(k_x,\omega)$ for different temperatures. When the superconducting gap is established, the spectrum exhibits sharp coherence peaks, a hallmark of the superconducting phase. As the temperature is decreased, these coherence features are gradually diminished and finally vanish as $T\to T_{c2}$, at which point the spectral response becomes effectively indistinguishable from an insulating state. This behavior signals a crossover in fluctuation physics: thermal fluctuations are suppressed with decreasing $T$, while quantum fluctuations strengthen and become the primary mechanism destabilizing superconductivity, thereby favoring an insulating ground state.

\begin{figure}
  \centering
  \includegraphics[width=0.45\textwidth]{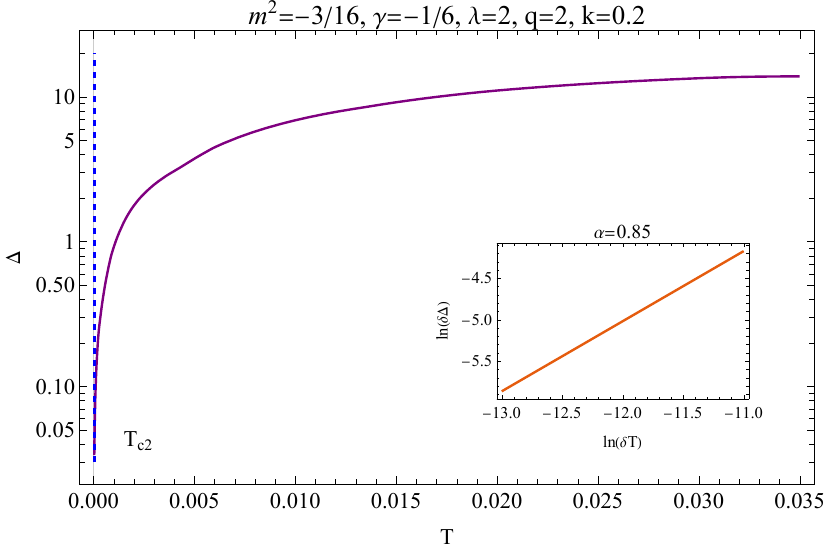}\qquad
  \includegraphics[width=0.45\textwidth]{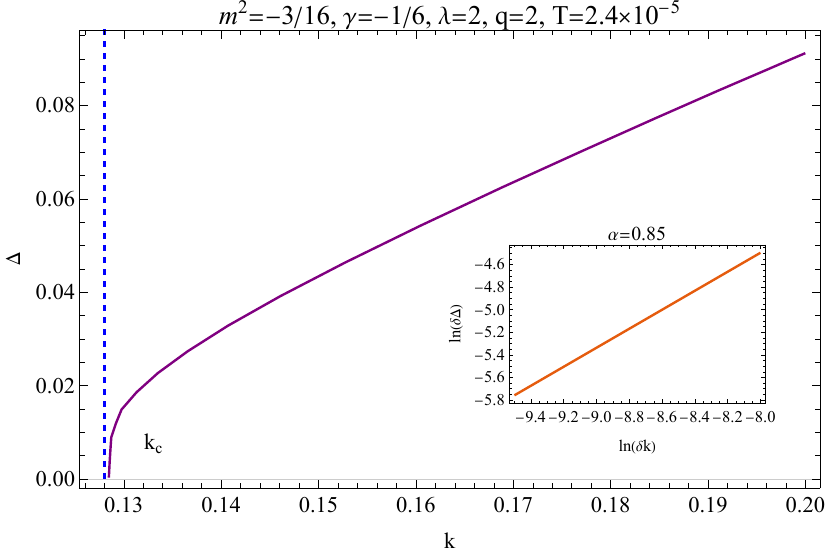}
  \caption{Scaling of the energy gap $\Delta$ with temperature $T$ and wave vector $k$. Left panel: $\Delta$ exhibits critical scaling at $T_{c2}$ with exponent $\alpha_T=0.85$. Right panel: $\Delta$ shows analogous scaling at $k_c=0.128$ with exponent $\alpha_k=0.85$.}
  \label{fig:scalingtem}
\end{figure}

Near the QCP, universality dictates that the gap exhibits scaling behavior $\Delta \sim (\xi-\xi_c)^{\alpha}$, where $\xi$ denotes the tuning parameter. Figure~\ref{fig:scalingtem} shows the corresponding critical behavior:
\begin{equation}
  \Delta\sim (T-T_{c2})^{\alpha_T},\qquad \Delta\sim (k-k_c)^{\alpha_k},
\end{equation}
with $\alpha_T = \alpha_k \approx 0.85$, confirming the QPT character. In the quantum critical regime, the system is controlled by a scale-invariant fixed point, leading to universal scaling. Since the SIT occurs at finite temperature in both our model and experiments, the system is described by a thermal density matrix and should be treated as a mixed state. Investigations of quantum information signatures across the SIT should therefore employ mixed-state entanglement measures to disentangle quantum from thermal contributions.
\section{Holographic quantum information}\label{sec:qi}

Having established the SIT and its critical scaling from the energy gap, we now turn to holographic quantum information measures to further characterize the QPT. HEE has been widely used to diagnose critical scaling~\cite{ling2016holographic,ling2016characterization,cai2012holographicsit,huang2020mixed}. However, the quantum critical regime is realized at finite temperature, so the boundary state is mixed and HEE includes thermal contributions. To properly characterize quantum correlations near a QPT, we consider EWCS, which provides a more faithful probe of mixed-state entanglement~\cite{yang2023mixed,liu2021mixed,liu2023mixed}.
\subsection{Definitions of HEE and EWCS}
For a pure state $|\psi\rangle$ bipartitioned into subsystem $A$ and its complement $B$, the EE is defined via the reduced density matrix~\cite{osterloh2002scaling,amico2008entanglement}:
\begin{equation}
  \rho_A=\mathrm{Tr}_B\bigl(|\psi\rangle\langle\psi|\bigr),
\end{equation}
from which we can obtain the entanglement entropy
\begin{equation}
  S_A(|\psi\rangle)=-\text{Tr}[\rho_A\text{log}(\rho_A)].
\end{equation}
Additionally, the holographic duality for EE, which is known as HEE, provides a geometric prescription for entanglement in AdS/CFT \cite{ryu2006holographic,nishioka2009holographic,casini2011towards}. In this framework, the HEE associated with a boundary subregion $A$ (and its complement $B$) is determined by the area of the bulk minimal surface extending into the bulk. It can be expressed as
\begin{equation}
  S_A=\frac{\text{Area} (\gamma_A)}{4 G_N^{d+2}}.
\end{equation}
$\gamma_A$ denotes the bulk minimal surface anchored on the boundary subregion $A$ and extending into the bulk. In this paper, we only consider the HEE for an infinitely long strip  along the $y$-direction (see in Appendix~\ref{ap:qi}). However, for sufficiently large configuration, the geometry of HEE is probably dominanted by the infrared (IR) region in the bulk, which means HEE receives substantial contributions from thermal entropy, rendering it an inadequate diagnostic of quantum correlations in mixed-state system. This motivates the introduction of the mixed-state entanglement measure. Among them, the EWCS plays a central role and has been conjectured to be dual to reflected entropy, logarithmic negativity, and odd entropy \cite{dutta2021canonical,kudler2019entanglement,jokela2019notes,camargo2022balanced,vasli2023holographic}. By definition, EWCS is the area of the minimal cross-section that partitions the entanglement wedge, and it is given by
\begin{equation}
  E_w(\rho_{AB})=\underset{\Sigma_{AB}}{\text{min}}\left(\frac{\text{Area}(\Sigma_{AB})}{4G_N}\right).
\end{equation}
$\Sigma_{AB}$ denotes the minimal cross-section. In this paper, we restrict to the translationally invariant setup of an infinitely long strip along the $y$-direction. In Appendix~\ref{ap:qi}, we schematically depicts the EWCS for a bipartite boundary configuration $a\cup c$ separated by an intermediate region $b$. The entanglement wedge is the bulk domain bounded by the corresponding minimal surface anchored on the boundary subregion, it exists as a connected bulk region only when the wedge is connected.

\begin{figure}
  \centering
  \includegraphics[width=0.75\textwidth]{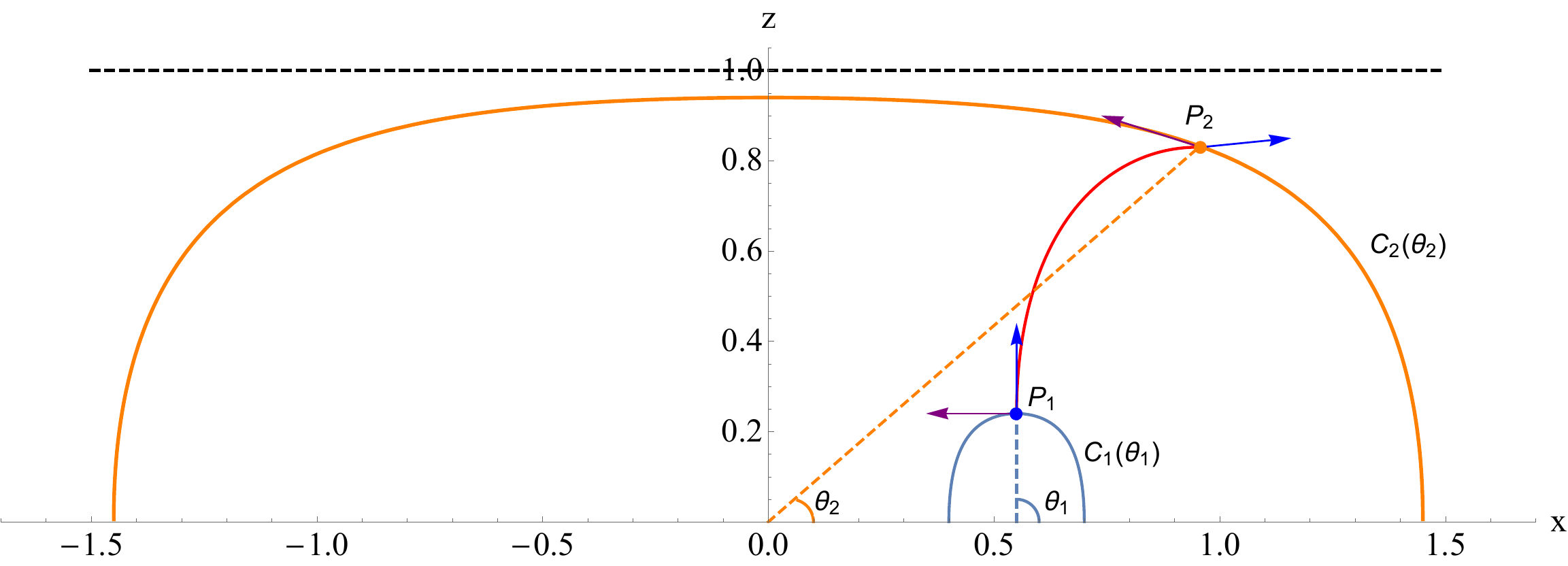}
  \caption{The schematic of the numerical solving the asymmetric EWCS.}
  \label{fig:algoewcs}
\end{figure}

The EWCS depends sensitively on the bipartite configuration. Asymmetric configurations are more physical but harder to compute than symmetric ones. The main obstacle is that locating the minimal entanglement-wedge cross-section typically requires scanning a two-dimensional parameter space, which is computationally difficult. Moreover, standard coordinate choices become singular near the AdS boundary, which can degrade numerical stability and precision. Motivated by these challenges, we develop an efficient and robust algorithm to evaluate the asymmetric EWCS \cite{liu2019entanglement}. The numerical calculation is schematically presented in figure~\ref{fig:algoewcs}. The background metric can be written as
\begin{equation}
  ds^2=g_{tt}dt^2+g_{zz}dz^2+g_{xx}dx^2+g_{yy}dy^2.
\end{equation}
We denote the minimal surfaces associated with the connected configuration by $C_1(\theta_1)$ and $C_2(\theta_2)$. Each surface is anchored to the cross-section slice at the points $P_1$ and $P_2$, respectively. Consequently, the area of the cross-section is
\begin{equation}
  A=\int_{C_{P_1}C_{P_2}}\sqrt{g_{xx}g_{yy}x'(x)^2+g_{zz}g_yy}dz.
\end{equation}
Therefore, the EOM of the local minimum cross-section can obtain from
\begin{equation}
  x'(z)^3\left (\frac{g_{xx}g_{yy}'}{2g_{yy}g_{zz}}+\frac{g_{xx}'}{2g_{zz}}\right )+x'(z)\left (\frac{g_{xx}'}{g_{xx}}+\frac{g_{yy}'}{2g_{yy}}-\frac{g_{zz}'}{2g_{zz}}\right )+x''(z)=0.
\end{equation}
Because the minimum cross-section is locally orthogonal to the boundary of the entanglement wedge, the relationship can be read as 
\begin{equation}\label{eq:orth}
  Q_1(\theta_1,\theta_2) \equiv\left.\frac{\langle \frac{\partial}{\partial z},\frac{\partial}{\partial \theta_1}\rangle}{\sqrt{\langle \frac{\partial}{\partial z},\frac{\partial}{\partial z}\rangle \langle\frac{\partial}{\partial \theta_1},\frac{\partial}{\partial \theta_1}\rangle}} \right |_{p_1}=0,\quad
  Q_2(\theta_1,\theta_2) \equiv\left.\frac{\langle \frac{\partial}{\partial z},\frac{\partial}{\partial \theta_1}\rangle}{\sqrt{\langle \frac{\partial}{\partial z},\frac{\partial}{\partial z}\rangle \langle\frac{\partial}{\partial \theta_2},\frac{\partial}{\partial \theta_2}\rangle}} \right |_{p_2} =0.
\end{equation}
Here, $\langle\cdot,\cdot\rangle$ denotes the inner product induced by the spacetime metric $g_{\mu\nu}$. The minimal cross-section is anchored on the minimal surface at $(\theta_1,\theta_2)$ once Eq. \eqref{eq:orth} is satisfied. To determine the endpoints $(P_1,P_2)$, we employ a Newton–Raphson iteration combined with a pseudospectral method. With this numerical algorithm, we then compute the asymmetric EWCS in the EMDA p-wave superconductor model, which serves as a quantitative probe of mixed-state entanglement in the system.

\subsection{Results for HEE and EWCS}
\begin{figure}
  \centering
  \includegraphics[width=0.45\textwidth]{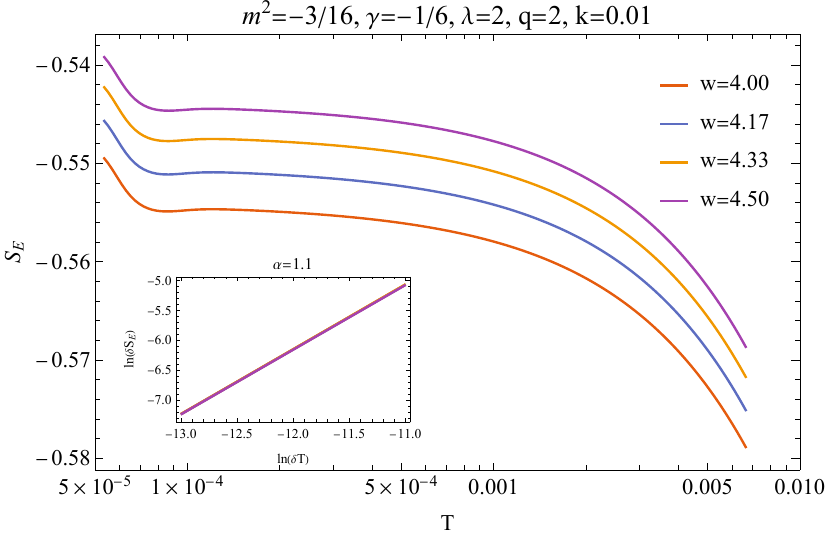}\qquad
  \includegraphics[width=0.45\textwidth]{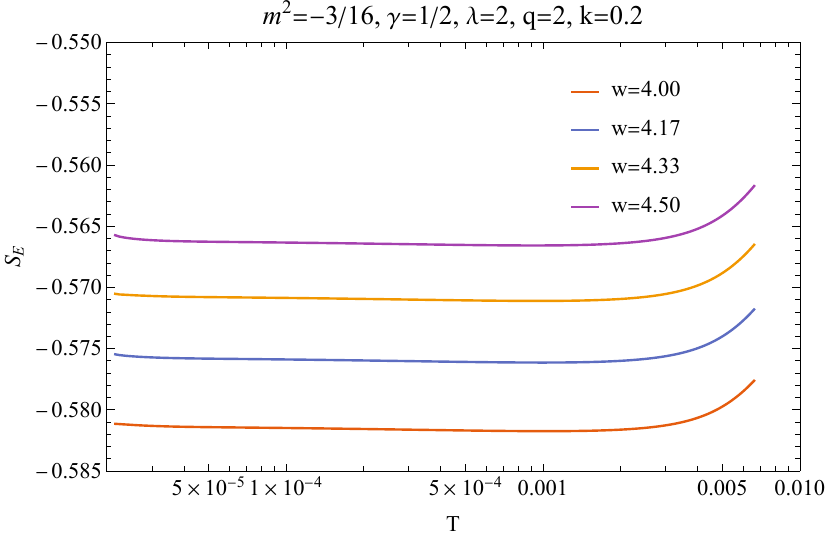}
  \caption{The behavior of the HEE $S_E$ as a function of temperature $T$ for several configurations $w$, the inset highlights the scaling behavior close to the QPTs. Since the UV-divergent contribution near the AdS boundary has been subtracted to render the HEE finite, the resulting renormalized $S_E$ can take negative values.}
  \label{fig:hee}
\end{figure}
\begin{figure}
  \centering
  \includegraphics[width=0.45\textwidth]{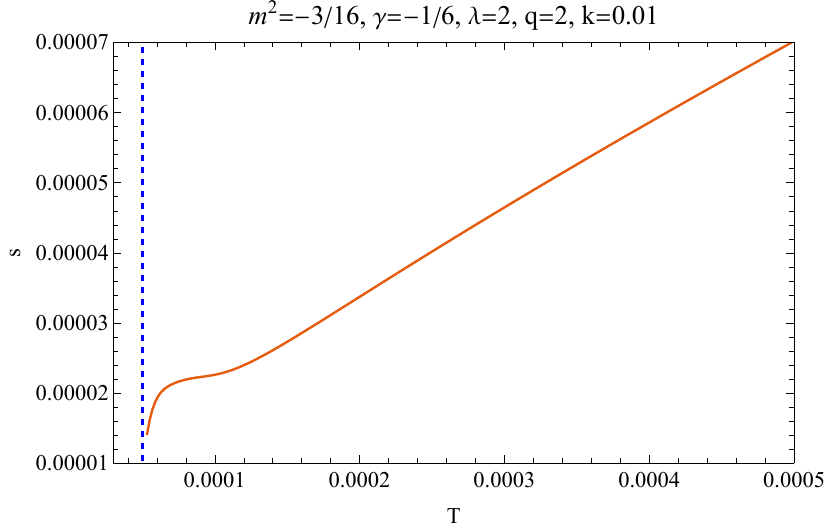}\qquad
  \includegraphics[width=0.45\textwidth]{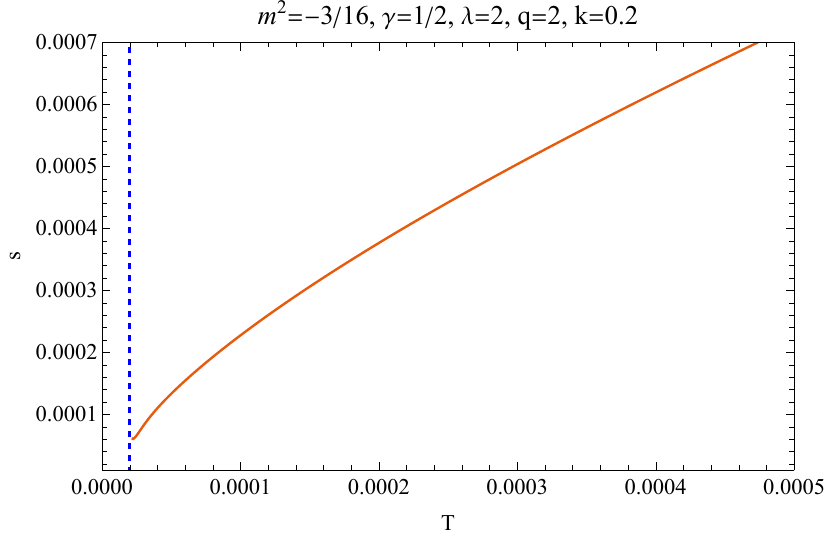}
  \caption{Entropy density $s$ versus temperature $T$. The blue dashed line marks the critical point.}
  \label{fig:entropy}
\end{figure}
\begin{figure}
  \centering
  \includegraphics[width=0.45\textwidth]{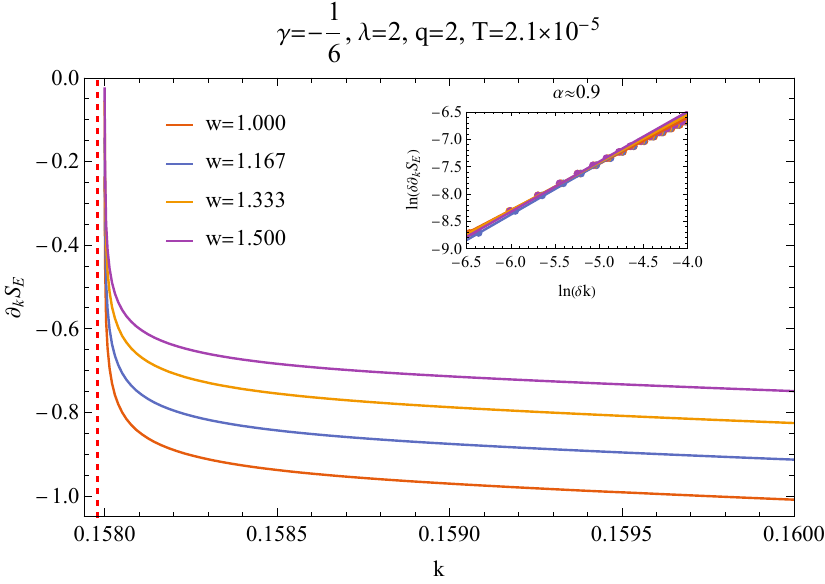}\qquad
  \includegraphics[width=0.45\textwidth]{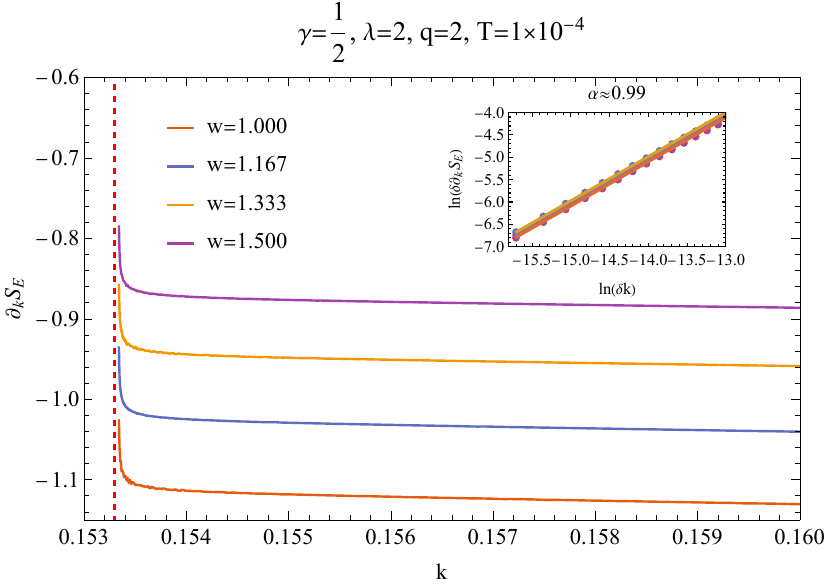}
  \caption{The behavior of the first derivative of HEE $\partial_k S_E$ as a function of wave vector $k$ for several configurations $w$, the inset highlights the scaling behavior close to the QPTs and the red lines represent the critical point. The inset figure is the scaling behavior of the $E_w$ and the red dashed lines represent the critical point.}
  \label{fig:heevsk}
\end{figure}

As illustrated in figure~\ref{fig:hee}, the HEE displays a pronounced temperature dependence near the QPT for several strip widths $w$. For $(\gamma, k)=(-1/6, 0.01)$, HEE develops scaling behavior in the vicinity of the critical point. In the non-critical region, HEE increases gradually as $T$ is lowered; upon crossing into the quantum critical regime, HEE undergoes a rapid upturn for large configurations, reflecting an abrupt enhancement of critical correlations. This motivates the scaling form:
\begin{equation}
  \delta S_E\sim (T-T_c)^\alpha,
\end{equation}
with $\alpha \approx 1.1$. However, for $(\gamma,k)=(1/2,0.2)$, scaling is absent. Since scaling is a phenomenological signal of phase transitions, its loss implies that HEE is not universally reliable for detecting QPTs in this model. This non-universality arises because, for large subsystems, the RT surface extends deep into the bulk and HEE becomes dominated by thermal entropy~\cite{ling2016holographic,ling2016characterization,yang2023mixed}. The entropy density,
\begin{equation}
  \tilde{s}=\frac{2\pi A}{\kappa^2}=\frac{2\pi\sqrt{V_1(z)V_2(z)}}{\kappa^2}\hat{V},
\end{equation}
where $A$ is the horizon area and $\hat{V}=\int dx\,dy$. We define the dimensionless entropy density $s=\frac{\kappa^{2}\tilde s}{2\pi\,\hat V\,\mu^{2}}$. Figure~\ref{fig:entropy} shows the behavior of $s$ versus $T$. For $(\gamma,k)=(-1/6,0.01)$, the entropy density exhibits scaling similar to HEE and decreases monotonically with temperature. By contrast, for $(\gamma,k)=(1/2,0.2)$, scaling disappears while $s$ still decreases with $T$. These observations indicate that, in large configurations, HEE scaling is strongly tied to thermal entropy, suggesting HEE is sensitive to the temperature. In contrast to temperature, the wave vector $k$ can be treated as an effective control parameter for quantum fluctuations in our setup. Figure~\ref{fig:heevsk} shows the behavior of the $\partial_k S_E$ near the QPTs. When $k$ approaches the critical point, we observe a pronounced scaling behavior, indicating enhanced critical sensitivity along the $k$ direction. Importantly, in stark contrast to the temperature dependence of HEE, the quantity $\partial_k S_E$ exhibits a clear capability to signal the QPTs for both parameter sets considered. This difference can be understood as follows, varying $T$ introduces substantial thermal entropy contributions to the HEE, which can mask the underlying quantum critical behavior and prevents HEE from serving as a universal QPT diagnostic. Conversely, tuning $k$ probes the critical quantum correlations more directly, making $\partial_k S_E$ a sharper indicator. To further reduce thermal effects and provide a more faithful characterization of the transition in mixed states, we therefore consider mixed-state entanglement measures, notably the EWCS.

\begin{figure}
  \centering
  \includegraphics[width=0.45\textwidth]{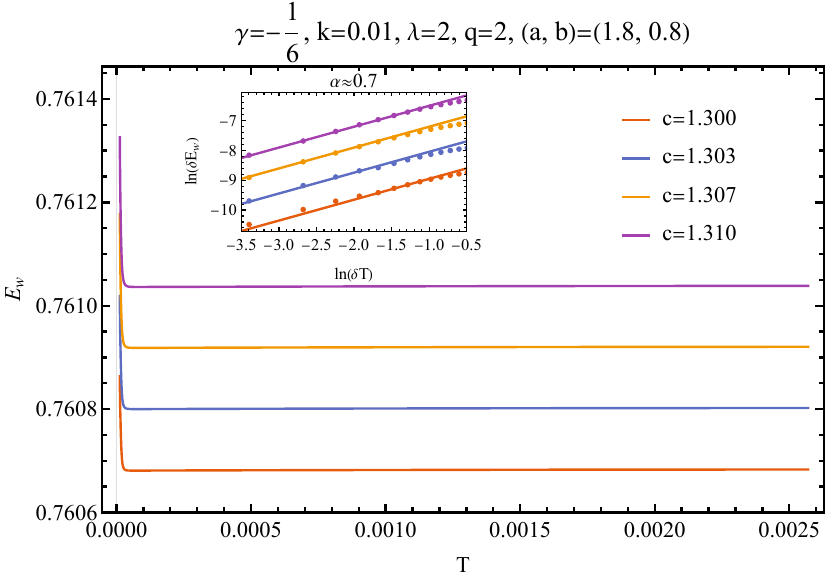}\qquad
  \includegraphics[width=0.45\textwidth]{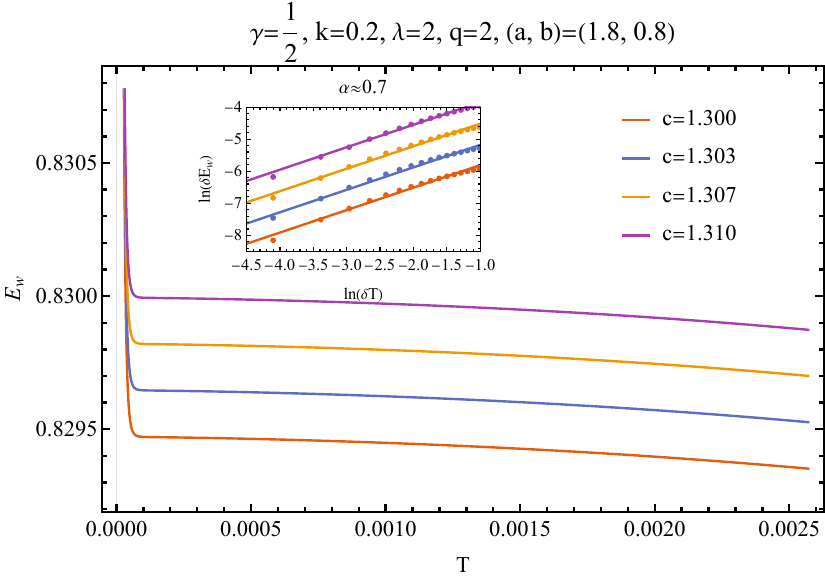}
  \caption{EWCS $E_w$ versus temperature $T$. Insets show the scaling behavior, yielding $\alpha\approx0.7$ for both parameter sets.}
  \label{fig:ewcs}
\end{figure}

\begin{figure}
  \centering
  \includegraphics[width=0.45\textwidth]{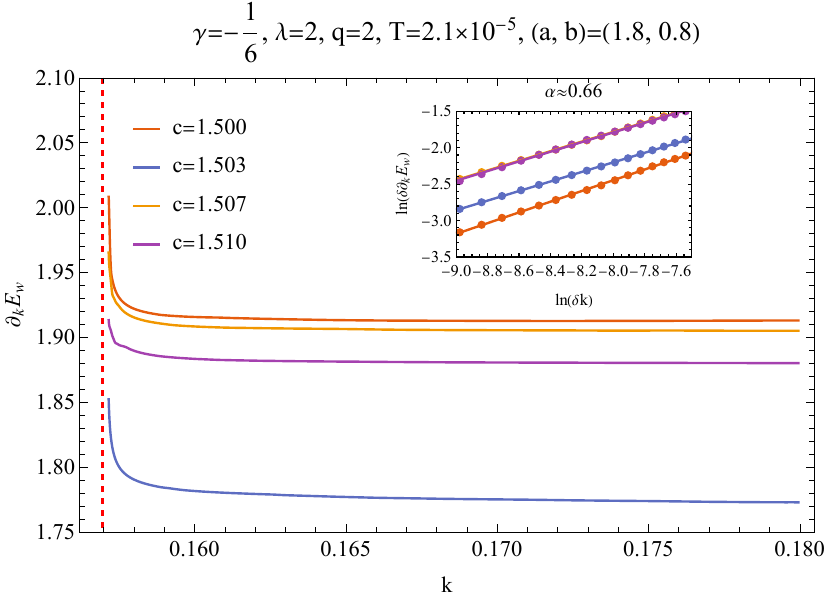}\qquad
  \includegraphics[width=0.45\textwidth]{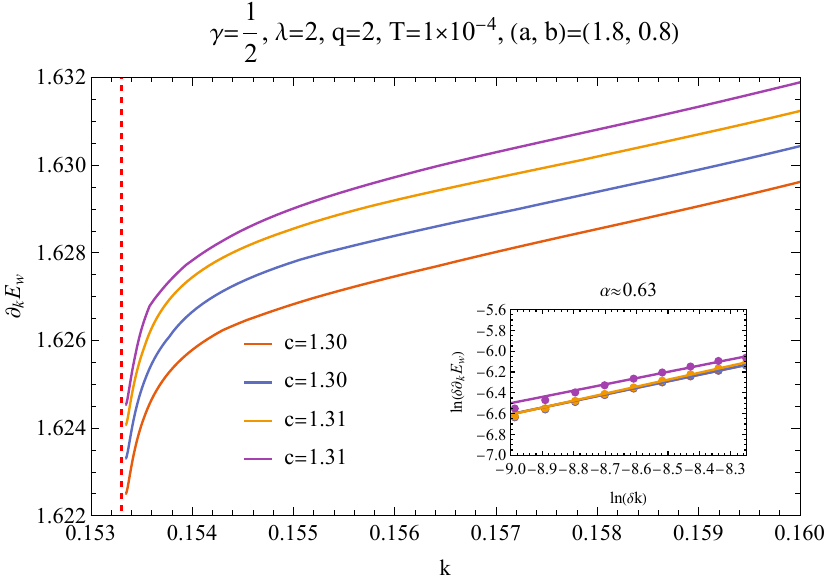}
  \caption{The behavior of the first derivative of EWCS $\partial_k E_w$ as a function of the wave vector $k$ for several different configurations. The inset figure is the scaling behavior of the $\partial_k E_w$ and the red dashed lines represent the critical point.}
  \label{fig:ewcsvsk}
\end{figure}

EWCS is a holographic probe of mixed-state entanglement that has been employed to diagnose phase transitions~\cite{huang2020mixed,yang2023mixed,liu2023mixed}. In figure~\ref{fig:ewcs}, we present the temperature dependence of EWCS. In the large-configuration regime, EWCS exhibits behavior qualitatively distinct from HEE and entropy density: it increases as $T$ decreases and displays clear scaling in the vicinity of the critical point with exponent $\alpha \approx 0.7$. Notably, even for $(\gamma,k)=(1/2,0.2)$, where neither HEE nor the entropy density signals the transition, EWCS still sharply characterizes the QPT. Figure~\ref{fig:ewcsvsk} shows that the derivative $\partial_k E_w$ also develops pronounced scaling near the QPT. This robust performance stems from the distinct geometric sensitivities of EWCS and HEE: while HEE is dominated by extremal surfaces anchored deep in the IR geometry, EWCS is determined by the minimal cross-section of the entanglement wedge and thus receives contributions from both the IR and intermediate bulk regions. By filtering thermal contributions to isolate quantum correlations, EWCS provides a more reliable probe of QPTs in mixed-state systems.

\section{Discussion}\label{sec:discu}
In this paper, we consider a holographic EMDA p-wave superconductor by introducing a vector order parameter into the EMDA framework. This extension leads to a notably richer phase structure. The system not only undergoes a superconducting phase transition—of either first or second order depending on parameters, but also exhibits an additional and unconventional transition within the superconducting regime: a SIT in the zero-temperature limit. We identify this transition as a novel quantum phase transition driven by the wave vector $k$. We treat $k$ as an effective control parameter for the axion field. A nonzero $k$ explicitly breaks translational symmetry and leads the deformation of the IR fixed point \cite{andrade2014simple,gouteraux2011generalized}. In the $T\to 0$ limit, the small-$k$ regime is driven toward insulating behaviour, and the system becomes more prone to exhibiting a SIT as $k$ is varied within this regime. The superconducting energy gap reflects this two-step structure, it firstly opens at the superconducting critical point, subsequently reaches its maximum, and then decreases upon approach to the QCP. Near to the superconductor–insulator critical point, the peak of the superconducting gap broadens and develops insulating-like features, and its scaling behavior as $T\to T_{c2}$ provides additional support for a quantum critical interpretation.

To further characterize the transition, we analyze holographic quantum-information measures by scanning both temperature and wave vector. In the temperature direction, the HEE displays strong variations near the QPTs for large entangling regions. Nevertheless, its temperature dependence is largely governed by thermal entropy, rendering it ineffective as a mixed-state entanglement diagnostic. In sharp contrast, the EWCS, an intrinsically mixed-state measure—exhibits a well-defined scaling regime in the vicinity of the QCP. When tuning the wave vector $k$, both HEE and EWCS develop clear scaling signals, implying that along the $k$ direction they can serve as efficient indicators of the QPTs. Importantly, across the quantum-critical regime the EWCS scaling is markedly more robust than that of the HEE, highlighting EWCS as a more reliable probe of criticality. This hierarchy can be understood geometrically, for large subsystems, the HEE is predominantly set by the IR region of the bulk geometry, whereas the EWCS receives contributions from both the IR and intermediate radial regions. The resulting enhanced sensitivity enables EWCS to encode a wider class of mixed-state correlations and thereby to provide a sharper diagnosis of quantum critical behavior.

Notably, the SIT emerges only in the EMDA p-wave superconductor, while the EMDA s-wave model exhibits no counterpart of this transition. This contrast can be naturally understood from symmetry considerations. The s-wave condensate is a scalar and, apart from the spontaneous breaking of the global $U(1)$, does not introduce additional symmetry. By contrast, the p-wave condensate is a vector order parameter and therefore admits a substantially larger set of symmetry-allowed couplings and anisotropies. For the specific choice of the condensed component $\rho_x$, the corresponding Ginzburg--Landau functional typically contains several competing minima related by the underlying symmetries \cite{leggett1975theoretical,sigrist1991phenomenological}. As a result, even small perturbations, such as crystalline anisotropy, external fields, or interaction-driven deformations can lift degeneracies among these minima and stabilize distinct ordered states, thereby enabling additional phase boundaries and transitions that are inaccessible in the s-wave case. Crucially, the EMDA geometry includes an axion profile that explicitly breaks translational symmetry along the $x$ direction. Since the condensate along $x$ ($\rho_x\neq 0$), this explicit anisotropy can couple efficiently to the order parameter, effectively acting as a control parameter that competes with superconducting and insulating behavior. This mechanism is reflected in figure~\ref{fig:qpt}, smaller values of $k$ make the EMDA background more susceptible to the insulating phase, and the SIT critical temperature increases accordingly. We therefore suggest that the SIT is not a general feature of holographic EMDA superconductors, but rather a consequence of the symmetry-breaking landscape of a vector (p-wave) condensate and explicit translation-symmetry breaking along the direction selected by the order parameter. This also explains why the SIT disappears in the EMDA s-wave superconductor model. A particularly interesting issue is how the SIT is modified when the system admits order parameters with two different orientations, i.e. when condensates along both the $x$ and $y$ directions are present simultaneously. The competition between these order parameters is expected to give rise to a more intricate phase diagram. Such a study would necessarily constitute a new model and is therefore beyond the scope of the present work, but represents a natural next step.

In the future, we will systematically determine the complete phase diagram of the EMDA p-wave system, and investigate possible inhomogeneous backgrounds that can arise in holographic p-wave superconductors and their impact on quantum chaos diagnostics. We also plan to generalize the setup to s-wave and d-wave condensates, where distinct symmetry-breaking patterns may yield further unconventional phases and entanglement structures.

\section*{Acknowledgments}
Peng Liu would like to thank Yun-Ha Zha, Yi-Er Liu and Bai Liu for their kind encouragement during this work. Zhe Yang appreciates Feng-Ying Deng's support and warm words of encouragement during this work. We are also very grateful to Wei-Jian Liang for their helpful discussion and suggestions. This work is supported by the Natural Science Foundation of China under Grant Nos. 12475054, 12275275, 12375055 and the Guangdong Basic and Applied Basic Research Foundation No. 2025A1515012063. Zhe Yang is supported by the Jiangsu Postgraduate Research and Practice Innovation Program under Grant No. KYCX25\_3922.

\appendix
\section{Equations of motion}\label{ap:eom}
The equations of motion (EOMs) follow from the action in Eq.~\eqref{eq:eom}. The dilaton field $\Psi$ satisfies
\begin{equation}
  \begin{aligned}
    &6\sinh(\Psi) - \frac{\gamma}{4}\cosh^{(\gamma-3)/3}(3\Psi)\sinh(3\Psi)\,F_{\mu\nu}F^{\mu\nu} \\
    &\quad + 3\nabla_\mu\nabla^\mu\Psi - 6\sinh(2\Psi)\nabla_\mu\chi\nabla^\mu\chi = 0.
  \end{aligned}
\end{equation}
The Maxwell field obeys
\begin{equation}
  \begin{aligned}
    &-2q^2 A_\mu\rho^\nu\rho^\dagger_\nu + q^2 A^\nu(\rho_\nu\rho^\dagger_\mu + \rho_\mu\rho^\dagger_\nu) \\
    &\quad - iq\rho^{\dagger\nu}\nabla_\mu\rho_\nu + iq\rho^\nu\nabla_\mu\rho^\dagger_\nu + iq\rho^{\dagger\nu}\nabla_\nu\rho_\mu + iq\gamma\rho^{\dagger\nu}\nabla_\nu\rho_\mu \\
    &\quad - iq\gamma\rho^\dagger_\mu\nabla_\nu\rho^\nu - iq\rho^\nu\nabla_\nu\rho^\dagger_\mu - iq\gamma\rho^\nu\nabla_\nu\rho^\dagger_\mu + iq\gamma\rho_\mu\nabla_\nu\rho^{\dagger\nu} \\
    &\quad - \gamma\cosh^{(\gamma-3)/3}(3\Psi)\sinh(3\Psi)\nabla_\mu A^\nu\nabla_\nu\Psi \\
    &\quad - \cosh^{\gamma/3}(3\Psi)\nabla_\nu\nabla_\mu A^\nu + \cosh^{\gamma/3}(3\Psi)\nabla_\nu\nabla^\nu A_\mu \\
    &\quad + \gamma\cosh^{(\gamma-3)/3}(3\Psi)\sinh(3\Psi)\nabla_\nu\Psi\nabla^\nu A_\mu = 0.
  \end{aligned}
\end{equation}
The axion field $\chi$ satisfies
\begin{equation}
  12\sinh(\Psi)\bigl[\sinh(\Psi)\nabla_\mu\nabla^\mu\chi + 2\cosh(\Psi)\nabla_\mu\Psi\nabla^\mu\chi\bigr] = 0,
\end{equation}
which is automatically satisfied by our ansatz $\chi = \hat{k}x$. The vector condensate field obeys
\begin{equation}
  \begin{aligned}
    &-m^2\rho_\mu - q^2 A_\nu A^\nu\rho_\mu + q^2 A_\mu A_\nu\rho^\nu - iq\gamma F_{\mu\nu}\rho^\nu \\
    &\quad + iq A_\nu\nabla_\mu\rho^\nu + iq\rho^\nu\nabla_\nu A_\mu - iq\rho_\mu\nabla_\nu A^\nu \\
    &\quad - iq A^\nu\nabla_\nu\rho_\mu + iq A_\mu\nabla_\nu\rho^\nu - \nabla_\nu\nabla_\mu\rho^\nu + \nabla_\nu\nabla^\nu\rho_\mu = 0.
  \end{aligned}
\end{equation}
The Einstein equations are
\begin{equation}
  \begin{aligned}
    &\mathcal{R}_{\mu\nu} - \tfrac{1}{2}g_{\mu\nu}\mathcal{R} - 3\cosh(\Psi)g_{\mu\nu} \\
    &\quad - \tfrac{1}{2}\cosh^{\gamma/3}(3\Psi)F_{\mu\sigma}F_\nu{}^\sigma + \tfrac{1}{8}\cosh^{\gamma/3}(3\Psi)F^2 g_{\mu\nu} \\
    &\quad - \tfrac{1}{4}\rho_{\nu\sigma}\rho^{\dagger\sigma}_\mu - \tfrac{1}{4}\rho_{\mu\sigma}\rho^{\dagger\sigma}_\nu + \tfrac{1}{4}g_{\mu\nu}\rho^{\sigma\tau}\rho^\dagger_{\sigma\tau} \\
    &\quad - \tfrac{1}{4}\rho_{\sigma\nu}\rho^{\dagger\sigma}_\mu - \tfrac{1}{4}\rho_{\sigma\mu}\rho^{\dagger\sigma}_\nu - \tfrac{1}{2}m^2\rho_\nu\rho^\dagger_\mu - \tfrac{1}{2}m^2 g_{\mu\nu}\rho^\sigma\rho^\dagger_\sigma \\
    &\quad - \tfrac{1}{2}iq\gamma F_{\nu\sigma}\rho^\sigma\rho^\dagger_\mu + \tfrac{1}{2}iq\gamma F_{\nu\sigma}\rho_\mu\rho^{\dagger\sigma} + \tfrac{1}{2}iq\gamma F_{\mu\sigma}\rho_\nu\rho^{\dagger\sigma} \\
    &\quad - \tfrac{1}{2}iq\gamma F_{\sigma\tau}g_{\mu\nu}\rho^\sigma\rho^{\dagger\tau} - 6\sinh^2(\Psi)\nabla_\mu\chi\nabla_\nu\chi + 3g_{\mu\nu}\sinh^2(\Psi)\nabla_\sigma\chi\nabla^\sigma\chi \\
    &\quad - \tfrac{3}{2}\nabla_\mu\Psi\nabla_\nu\Psi + \tfrac{3}{4}g_{\mu\nu}\nabla_\sigma\Psi\nabla^\sigma\Psi = 0.
  \end{aligned}
\end{equation}

\section{The schematic of the holographic quantum information}\label{ap:qi}
\begin{figure}
  \centering
  \begin{tikzpicture}[scale=1]
    \node [above right] at (0,0) {\includegraphics[width=8cm]{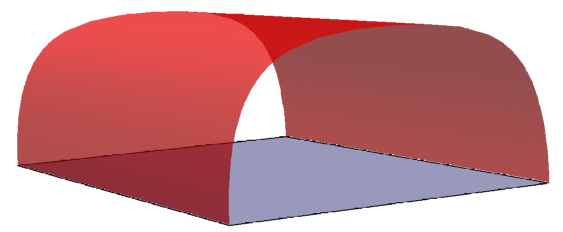}};
    \draw [right,->,thick] (3.4, 0.2) -- (5.25, 0.5) node[below] {$x$};
    \draw [right,->,thick] (3.4, 0.2) -- (1.85, 0.7) node[below] {$y$};
    \draw [right,->,thick] (3.4, 0.2) -- (3.4, 3.425) node[above] {$z$};
  \end{tikzpicture}
  \begin{tikzpicture}[scale=1]
    \node [above right] at (0,0) {\includegraphics[width=7.3cm]{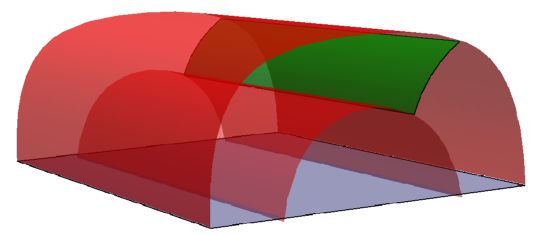}};
    \draw [right,->,thick] (2.94, 0.221) -- (4.55, 0.45) node[below] {$x$};
    \draw [right,->,thick] (2.94, 0.221) -- (1.45, 0.75) node[below] {$y$};
    \draw [right,->,thick] (2.94, 0.221) -- (3.0, 3.425) node[above] {$z$};
  \end{tikzpicture}
  \caption{Schematic illustration of the geometric prescriptions for HEE and EWCS. Left panel: The red surface is the minimal surface extending into the bulk and anchored on the boundary of the entangling region, according to the RT formula, the area determines the HEE. Right panel: The green surface is the minimal cross-section within the entanglement wedge, the EWCS obtained by minimizing the area among surfaces that separate the entanglement wedge into the corresponding boundary subregions.}
  \label{fig:schematic}
\end{figure}

In figure~\ref{fig:schematic}, we schematically summarize the geometric constructions underlying HEE and EWCS. We focus exclusively on an infinite strip geometry, translationally invariant along the $y$-direction. The configuration consists of two disjoint entangling subregions, $a$ and $c$, with their separation controlled by the intervening region $b$.

\end{document}